# FRONT MATTER

## Title
- Full titles: Two-dimensional superconductivity of the Ca-intercalated graphene on SiC: vital role of the interface between monolayer graphene and the substrate
- Short title: 2D superconductivity in Ca-intercalated graphene

## Authors

Haruko Toyama,[1*] Ryota Akiyama,[1*] Satoru Ichinokura,[2] Mizuki Hashizume,[2] Takushi Iimori,[3] Yukihiro Endo,[1¶] Rei Hobara,[1] Tomohiro Matsui,[1†] Kentaro Horii,[2] Shunsuke Sato,[1] Toru Hirahara,[2] Fumio Komori,[3#] Shuji Hasegawa[1]

## Affiliations
[1]Department of Physics, The University of Tokyo, Tokyo 113-0033, Japan [2]Department of Physics, Tokyo Institution of Technology, Tokyo 152-8551, Japan [3]The Institute for Solid State Physics, The University of Tokyo, Chiba 277-8581, Japan

*Corresponding author E-mail:
h.toyama@surface.phys.s.u-tokyo.ac.jp
akiyama@surface.phys.s.u-tokyo.ac.jp

Present address: [¶]NTT Basic Research Laboratories, NTT Corporation, 3-1 Morinosato-Wakamiya, Atsugi, Kanagawa 243-0198, Japan, [†]Advanced Research Laboratory, Anritsu corporation, 5-1-1 Onna, Atsugi, Kanagawa 243-8555, Japan, [#]Institute of Industrial Science, The University of Tokyo, Tokyo 153-8505, Japan

## Abstract

Ca-intercalation has opened a way for superconductivity in graphene on SiC. However, the atomic and electronic structures being critical for superconductivity are still under discussion. We find the essential role of the interface between monolayer graphene and the SiC substrate for superconductivity. In the Ca-intercalation process, at the interface a carbon layer terminating SiC changes to graphene by Ca-termination of SiC (monolayer graphene becomes bilayer) with inducing more carriers than a free-standing model. Then, Ca is intercalated in-between graphene layers, which shows superconductivity with the updated critical temperature ($T_C$) of up to 5.7 K. In addition, the relation between $T_C$ and the normal-state conductivity is unusual, "dome-shape". These findings are beyond the simple $C_6CaC_6$ model in which s-wave BCS superconductivity is theoretically predicted. This work proposes a general picture of the intercalation-induced superconductivity in graphene on SiC, and shed the light on the potential of superconductivity induced by other intercalants.

## Teaser
Intercalated-Ca terminates SiC and goes in-between bilayer graphene, which makes superconductivity beyond a simple BCS model.

# MAIN TEXT

**Introduction**

Graphene, a well-known atomic-layer two-dimensional (2D) carbon sheet consisting of a honeycomb lattice, has been attracting continuous research interests over decades due to its characteristic properties such as massless Dirac electrons, very high carrier mobility, and valley degree-of-freedom, together with a great advantage of the robustness under atmospheric conditions. Superconductivity is one of the important targets in intensive graphene research (1). In 2016, our group reported that the Ca-intercalated bilayer graphene grown on a silicon carbide (SiC(0001)) substrate shows the superconducting transition at $T_C$ = 4 K by transport measurements (2). In the early stages, the superconductivity was naively believed to be induced by the interlayer band (ILB) which is a free-electron-like (FE-like) parabolic band localized between graphene layers (3,4) from the analogy with the superconductivity in bulk graphite intercalation compounds (GIC) (5-8). However, in graphene, we need to take into account the importance of the ILB together with the $\pi^*/\pi$ bands which is proposed by calculation (9-11) and experimentally indicated in the Li-intercalation graphene with angle-resolved photoemission spectroscopy (ARPES) (12). Actually, as an alternative origin for superconductivity in graphene, recent theoretical studies mention flat $\pi^*$ bands at the M point accompanied by the van Hove singularity (vHs) which can induce the unconventional superconductivity by electron doping (13-16). However, little information about the origin of the superconductivity in the Ca-intercalated graphene has been offered so far.

Besides, the atomic structure of Ca-intercalated graphene on SiC substrate has not yet been fully clarified. Although Ca atoms were naively believed to be intercalated between graphene layers, it was recently reported from results of total-reflection high-energy positron diffraction (TRHEPD) (17) that Ca atoms are inserted not in-between the bilayer graphene but in the interface between graphene and SiC substrate. Specifically, they suggested that Ca atoms are located between the epitaxial graphene and a buffer layer that is a carbon layer covalently bonded with the topmost Si atoms on the surface of SiC(0001) substrate. This encourages us not only to reconsider the model of in-between-intercalation, but also to raise a question of whether superconductivity can be induced by the Ca-intercalations even in monolayer graphene on SiC. In addition, it has been reported that many other elements such as Li (18-22), Ca (23), Pb (24), Yb (25), and In (26,27) are intercalated between the buffer layer and the SiC surface. Those atoms break the C-Si bonds and terminate the SiC surface, transforming the buffer layer into a free-standing graphene layer. Therefore, for the atomic structure model of the Ca-intercalated graphene, we have to take into account the structural change around the interface between graphene and SiC substrate, which will lead to a better understanding of the electronic structures and the nature of the superconductivity.

In this report, we present that the interface between the epitaxial graphene and the substrate plays an important role in superconductivity in the Ca-intercalated graphene on SiC(0001). By tracking the evolutions in electron diffraction patterns, band dispersions, and transport properties during the intercalation process, we have firmly revealed that the superconductivity occurs in the Ca-intercalated bilayer graphene that is generated from a "monolayer" graphene by transforming the buffer layer at the interface to a graphene layer below the first graphene layer. It should be noticed here that the present bilayer graphene is directly on the SiC substrate while the bilayer graphene in our previous research in 2016 (2) has the buffer carbon layer beneath the bilayer on the SiC substrate. We have also found that $T_C$ as well as the normal-state conductivity change with small differences in the sample preparation, resulting in a dome-shaped dependence in a phase diagram of $T_C$ vs. the normal-state conductivity. The highest $T_C$ = 5.7 K at the optimized condition for the bilayer graphene where the large Fermi surface by the heavily-doped double π* bands as well as the ILB appears. The dome-shaped phase diagram suggests a possibility of unconventional superconductivity. A feature of the two-dimensional superconductivity, Berezinskii-Kosterlitz-Thouless (BKT) transition, is observed. The substantial carrier density of more than $10^{15}$ cm$^{-2}$ cannot be explained solely by the model of free-standing Ca-intercalated bilayer graphene ($C_6CaC_6$), which implies the contribution from additional charge transfer from the Ca-terminated surface of the SiC substrate. In this system, the key for changing atomic/electronic properties is the structural transformation at the interface between graphene and the SiC substrate, which enables the superconductivity even with monolayer graphene on SiC.

**Results**

The pristine graphene was obtained by the direct heating of an *n*-type SiC(0001) wafer in the Ar atmosphere (28). In this study, the fabrication process of the Ca-intercalated graphene was monitored by complementary analyses of the surface structure by reflection-high-energy electron diffraction (RHEED) in Fig. 1(a-d), the electronic band dispersions by ARPES in Fig. 1(e-h), and the microscopic surface morphology by scanning tunneling microscopy (STM) in Fig. 1(i). In Fig. 1(a), the RHEED pattern of the pristine graphene before the metal deposition shows bulk SiC 1×1 spots (white arrows) and 1×1 streaks of the topmost graphene layer (red arrows). The (6√3×6√3)R30º spots with respect to the SiC 1×1 encircled by a gray line originate from the buffer layer located at the interface between graphene and the SiC substrate. The sample was confirmed to be monolayer because a single Dirac cone at the K point in the Brillouin zone (BZ) of graphene was observed by ARPES (Fig. 1(e)). It is quite difficult to grow perfect monolayer graphene exclusively so that some regions of bilayer graphene also inevitably exist on the surface.

In this work, we employed an atom-replacement method from Li to Ca for the Ca intercalation (3). When Li is deposited on the pristine graphene at room temperature (RT), (6√3×6√3)R30º spots of the buffer layer gradually disappear (see Fig. S2(b)) followed by the appearance of (√3×√3)R30º spots indicated by green arrows in Fig. 1(b). The attenuation and disappearance of (6√3×6√3)R30º spots indicate that Li atoms are inserted below the buffer layer, resulting in the Li-termination of the SiC surface and decoupling the buffer layer from the SiC surface. This means that the buffer layer changes into a quasi-free-standing graphene layer, forming bilayer graphene on top of the Li-terminated SiC surface (18-20,22). Thanks to this change from the monolayer graphene to the bilayer one, excess Li atoms can go into in-between two graphene layers, too (29). This is confirmed also by our ARPES results; the double Dirac cones are observed as evidence of the bilayer graphene formation after the Li-deposition (Fig. 1(f)). Here we define the π* band of the inner and outer Dirac cones as π*$_1$ and π*$_2$, respectively. Note that the Dirac cones which are originally located at the K points, are also observable at the Γ point as replica bands because the second Γ' point in the BZ of (√3×√3)R30º is equivalent to the K point in 1×1-BZ of graphene.

Two parabolic bands with convex upwards whose tops are at ~ -0.7 eV and ~ -1.2 eV (indicated by white dotted lines in Fig. 1(f)) are the bulk valence bands of SiC substrate, which are shifted upward compared with those before Li deposition due to the change of the surface band bending by the Li-termination. The magnitude of the band bending is consistent with the reported core level shift, interpreted to be originated from the dipole layer at the interface between graphene and SiC by the Li termination (20). This result also supports the decoupling of the buffer layer from the SiC substrate after the Li-termination.

Fig. 1(c) shows the RHEED pattern after the Ca-deposition on the Li-intercalated graphene at ~ 180°C. There appear new spots (blue arrows) at the slightly different positions from those of ($\sqrt{3}\times\sqrt{3}$)R30º spots. These new spots originate from three-dimensional (3D) bulk Ca(111) islands on the topmost surface of the Li-intercalated graphene. Here, Li atoms should remain in-between graphene layers as well as at the surface of SiC because the substrate temperature during the Ca-deposition is kept lower than ~280°C at which Li atoms start desorbing (17). In fact, the band dispersion of the Ca-on-top-sample (Fig. 1(g)) is quite similar to that of the Li-intercalated graphene before Ca-deposition (Fig. 1(f)); the double Dirac cones originating from the bilayer graphene are still observed with almost the same binding energy at the Γ' point, which indicates that ($\sqrt{3}\times\sqrt{3}$)R30º periodicity remains due to the intercalated Li between the graphene layers (green arrows in Fig. 1(c)). Moreover, carriers are found not to be doped since the bands do not shift in energy even after Ca deposition on the surface (Fig. 1(g)).

After annealing the Ca-deposited sample up to ~ 300°C, the structures drastically changed. First, the 3D spots from the bulk Ca(111) disappeared, suggesting the desorption of excess Ca atoms, and clearer ($\sqrt{3}\times\sqrt{3}$)R30º spots (orange arrows) appeared (Fig. 1(d)). The strong ($\sqrt{3}\times\sqrt{3}$)R30º spots should originate from intercalated Ca atoms in-between graphene layers because the annealing temperature is too high for the intercalated Li atoms to stay in-between the graphene layers so that the Li atoms between the graphene layers are replaced by Ca atoms, as proposed by Kanetani *et al.* (3). This is also supported by the positron diffraction analysis (17). As shown in Fig. 1(i), the ($\sqrt{3}\times\sqrt{3}$)R30º structure by the intercalated Ca is clearly observed also by STM together with 1×1 unit cells of graphene, which confirms the high quality of the sample in the atomic scale (see Fig. S2(d-f) for the details).

Fig. 1(h) shows the band dispersion of the Ca-intercalated sample. The double Dirac cones of the bilayer graphene shift downward in energy compared with Fig. 1(f), meaning further electron-doping. This is definite evidence that the bilayer structure is preserved after the annealing and the Li-Ca replacement occurs. Thus, these ARPES results indicate that the Ca-intercalation makes the monolayer graphene + buffer layer into quasi-free-standing bilayer graphene. When the sample is annealed at much higher temperature (> ~ 400°C), the intercalated-Ca-atoms finally desorb from the graphene layers, whereas the atoms terminating the SiC surface still remain, resulting in the quasi-free-standing bilayer graphene with no in-between intercalated atoms. The details about this de-intercalation process are described later in the results of transport measurement.

The schematic models of the cross-sectional structures at each step of intercalation into graphene are summarized in Figs. 2(a-e). The pristine state has monolayer graphene (black balls) and an underlying buffer layer (grey balls) bonding to the SiC substrate (Fig. 2(a)). When Li is deposited on it, Li atoms (green balls) are inserted beneath the buffer layer with breaking bonds there and terminate the SiC surface (Fig. 2(b)). As a result, the buffer layer becomes quasi-free-standing graphene, and the total number of graphene layers increases now from one to two (bilayer graphene). After finishing the Li-termination, excess Li atoms are intercalated between the graphene layers (Fig. 2(c)). Then, when Ca atoms (yellow balls) are just deposited on it at ~ 160°C, 3D Ca(111) islands are formed on the topmost surface (Fig. 2(d)). Finally, after the sample is heated up to a high enough temperature, Ca atoms are intercalated between the graphene layers by replacing Li atoms, as shown in Fig. 2(e). As described later, Ca replaces with Li atoms terminating SiC, and simultaneously Ca terminates partially unterminated bonds on the SiC surface.

Figures 2(f-i) show the results of transport measurements. The temperature dependences of the sheet resistance $R(T)$ of the sample at each step of the intercalation mentioned above are summarized in Fig. 2(f). First, the Li-intercalated graphene (blue line)

(corresponding to Fig. 2(c)) shows a metallic behavior with no superconducting transition down to ~ 0.8 K, which is consistent with the previous work on the Li-intercalated bilayer graphene + buffer layer (30). After the Ca-deposition on top (green line) (corresponding to Fig. 2(d)), the non-superconducting behavior does not change while the resistance slightly decreases from the Li-intercalated one. However, the superconducting transition occurs when Ca atoms are intercalated into the graphene layers by annealing (red line) (corresponding to Fig. 2(e)). This result indicates that even if graphene is initially monolayer, the superconducting Ca-intercalated bilayer graphene can be fabricated because the buffer layer is lifted by the preceding intercalation of Li underneath it. The superconducting transition disappears when the Ca-intercalated sample is heated up to ~ 400°C at which the intercalated Ca atoms desorb from the graphene layers (yellow line). Therefore, we conclude that the superconductivity is obviously induced by the Ca-intercalation in-between the graphene layers. This is also supported by the ARPES results in Fig. 1 that the Li-intercalated and the Ca-deposited samples (Fig. 2(d)) have almost the same band structures as the Li-intercalated one (Fig. 2(c), whereas the Ca-intercalated sample (Fig. 2(e)) has Dirac cones which are significantly shifted downwards in energy. Note that the increase of the normal-state-resistance in yellow data Fig. 2(f) is caused by the defects induced by the Ca-intercalation and desorption processes, as indicated in previous work (30).

As seen in Fig. 2(f), the Ca-intercalated graphene shows two shoulders in the $R(T)$ curve at 5.7 K and 4.2 K, which indicates that two kinds of superconducting transitions having different critical temperatures coexist on the surface. We speculate that this two-step transition originates from the coexisting monolayer and bilayer regions, which have different carrier densities per graphene layer.

Figures 2(g) and (h) show the sheet resistance as a functions of the magnetic field and temperature, respectively. In all measurements, the magnetic field was applied perpendicular to the sample plane. In Fig. 2(g), the critical field for breaking the superconductivity is lower with increasing the temperature, and in Fig. 2(h) one can see the superconducting transition temperature decreases with increasing the magnetic field. Here we define the upper critical magnetic field perpendicular to the sample plane $H_{c2}$ and $T_C$ as the values where the sheet resistance becomes a half of the normal-state resistance 96.6 Ω. In Fig. 2(i), $\mu_0 H_{c2}$ is plotted as a function of temperature. The values of blue and red dots were estimated from the data of Figs. 2(g) and 2(h), respectively. The linear relation of $H_{c2}$ with temperature represented by the dashed black line is a typical property in two-dimensional (2D) superconductors which can be described by Ginzburg-Landau (GL) theory as follows:

$$\mu_0 H_{c2}(T) = \frac{\phi_0}{2\pi \xi_{GL}(0)^2}\left(1 - \frac{T}{T_c}\right) \qquad (1)$$

where $\phi_0$ is the flux quantum and $\xi_{GL}(0)$ is the GL coherence length, respectively. From the GL fitting results, the GL coherence length and the upper critical field are calculated as $\xi_{GL}(0) = 35.2$ nm and $H_{c2}(0) = 0.27$ T, respectively, which are similar to those of the bulk Ca-GIC, $\xi_{GL}(0) = 29-36$ nm and $H_{c2}(0) = 0.25-0.50$ T (31-34). The obtained values of $\xi_{GL}(0)$ and $H_{c2}(0)$ are also comparable to those of other reported 2D superconductors (*e.g.*, $\xi_{GL}(0) = 22.2$ nm and $H_{c2}(0) = 0.67$ T in (Tl,Pb)/Si(111) (35)), which suggests the reasonability of GL fitting in our case.

One of the characteristic features of the 2D superconductivity is the BKT transition. Directly below $T_C$, the resistivity shows a finite value due to fluctuation of vortex-antivortex creation/annihilation which prevents the long-range phase coherence to produce a non-zero resistance state. By further cooling below the BKT transition temperature ($T_{BKT}$), the zero-resistance is attained. With the theory of Halperin-Nelson on the BKT transition (36), we fitted the measured $R(T)$ data of the superconducting Ca-intercalated graphene as shown in Fig. 3(a) by the following formula;

$$T - T_{BKT} \propto \left(\frac{d \ln R(T)}{dT}\right)^{-2/3}. \tag{2}$$

The fitting line is indicated by a black dashed curve. In Fig. 3(b), to see the BKT temperature in detail, the data are replotted with the vertical axis of $(d \ln(R(T))/dT)^{-2/3}$, and the straight fitting line provides $T_{BKT}$ to be 2.5 K as indicated by a black arrow.

The BKT transition can be seen also in the measured current-voltage (*I-V*) curves as shown in Fig. 3(c) with a double-logarithmic scale. We fitted the *I-V* curves in the transition region between the zero-resistance state and the normal state by $V \propto I^\alpha$ which are indicated by black dashed lines. The obtained power $\alpha$ is shown as a function of temperature in the inset. When the temperature increases, the value of $\alpha$ gradually decreases and finally approaches unity, *i.e.*, the ohmic behavior of the normal state above $T_C$. Theoretically, $\alpha$ becomes 3 at $T_{BKT}$, which is ~ 2.1 K from the graph, in good agreement with the obtained $T_{BKT}$ = 2.5 K in Fig. 3(b) with the margin of error (see Supplementary Text about the error). This first observation of the BKT transition in the Ca-intercalated graphene clearly supports the two-dimensionality of the superconductivity. Compared with 2D superconductors showing the steep transition between $\alpha$ = 1 and 3 in the $\alpha$-$T$ graph (35), in the present Ca-intercalated graphene the transition of $\alpha$ is gradual as well as the transition of superconductivity as reported in previous cases (37-39). This broadening in the transition of $\alpha$ is proposed to occur in the inhomogeneous system (40). Indeed, we suppose that our present samples are inhomogeneous more or less due to the inevitable coexistence of monolayer and bilayer graphene regions on the surface as indicated by the two-step transition in Fig. 2(f).

Next, we focus on the property of $T_C$. Figure 3(d) shows $T_C$ as a function of the conductivity at the normal state. Data points are for samples with different amounts of the deposited Li and Ca atoms, annealing temperatures, and the number of intercalation cycles defined as a series of Li-intercalation, Ca-deposition, and annealing (see Fig. S3(a) for the original data of $R(T)$ for the respective samples). The normal-state conductivity and $T_C$ are different from sample to sample depending on the details of sample preparation conditions. Red and blue dots indicate $T_C^{onset}$ where $R(T)$ starts to decrease with decreasing temperature and $T_C^{zero}$ where $R(T)$ becomes zero, respectively. Thus, the area shaded by the pale red color is the superconducting region and the outside of it is the normal metal region. Both serieses of $T_C^{onset}$ and $T_C^{zero}$ are found to be dome-shaped with taking a maximum at ~ 10 mS (~ 100 Ω). The maximum value of $T_C$ is 5.7 K, which updates the reported value (2).

The samples of less than 10 mS may simply mean insufficient intercalation. This is also seen in Fig. 3(e); with the intercalation cycles less than ~ 3 times, $T_C$ increases as the number of cycles increases by expanding the intercalated area (see Fig. S2(d)) and resultant increasing the carrier density. On the other hand, the suppression of $T_C$ over 10 mS in Fig. 3(d) indicates that $T_C$ is not simply enhanced with the increase in carrier density. This is more clearly seen in Fig. S3(b-d); the increase of the normal-state conductivity and the suppression of $T_C$ occur simultaneously by additional deposition of a small amount of Li which provides more electrons with almost no change in the surface structures. With these results, it is suggested that the suppression of $T_C$ over 10 mS is caused by the increase of the carrier density because the mobility is not expected to increase with the additional deposition of Ca or Li. When, furthermore, the intercalation cycles is more than ~ 3 times, $T_C^{zero}$ starts to decrease (Fig. 3(e)) while the normal-state conductivity increases due to more intercalation. Although we have no further evidence, these results do not seem in accordance with the BCS theory in which $T_C$ should increase with the density of states (carrier density) at the Fermi level $E_F$.

Next, we investigate changes in the shape, size, and carrier type at the Fermi surface. In heavily doped graphene, when $E_F$ exceeds the binding energy of vHs point ($E_{vHs}$) at M' point, the Fermi surface changes from electron pockets centered at the K point to large hole pockets centered at the Γ point because the vHs point is a saddle point in the band dispersion as schematically shown in Figs. 4(c-f). In order to estimate the carrier (electron) density, the Fermi surface mapping was made in the Li- and Ca-intercalated graphene samples as shown in Figs. 4(a) and (b), respectively. In the Fermi surface of the Li-intercalated graphene, the inner Dirac cone $\pi^*_1$ has a snowflake-like shape around the Γ point which is the overlapped replica bands reflecting electron pockets at each K point. The size of the Fermi surface becomes ~ 48 times larger than that of the pristine graphene since the graphene layers are heavily doped by the intercalated Li atoms. On the other hand, the outer $\pi^*_2$ at the M' point has a short line-shaped Fermi surface which is a part of the replica band of the vHs of $\pi^*$ band originated from the M point as shown in Fig. 4(a) corresponding to the schematic of Fig. 4(e). This indicates that $E_F$ is just tuned to $E_{vHs}$ in the Li-intercalated graphene. From this Fermi surface mapping, the electron density in a BZ is calculated to be $n_{Li,\pi^*1} = 1.8 \times 10^{14}$ cm$^{-2}$ for $\pi^*_1$ and $n_{Li,\pi^*2} = 3.6 \times 10^{14}$ cm$^{-2}$ for $\pi^*_2$, resulting in totally $n_{Li,\pi^*} = n_{Li,\pi^*1} + n_{Li,\pi^*2} = 5.3 \times 10^{14}$ cm$^{-2}$ in the $\pi^*$ bands. On the other hand, as shown in Fig. 4(b) for the Ca-intercalated sample, the situation is quite different: It is obvious that the size of the snowflake-like $\pi^*_1$ band around the Γ point is larger than that of the Li-intercalated graphene, meaning a higher doping of electrons by the intercalated Ca. Actually, the Dirac point of $\pi^*_1$ shifts by ~ 0.3 eV downwards in energy than that of the Li-intercalated case as shown in Figs. 1(f,h). Sharp-pointed edges of the snowflake extend from the Γ towards the M' point but do not reach to M'. This shape of the band around the M' point corresponds to Fig. 4(d) where the vHs of $\pi^*_1$ band is located above the $E_F$, suggesting that $\pi^*_1$ band is an electron pocket. The Fermi contour of $\pi^*_2$ is also seen around the M' point, but it splits into two lines surrounding the M' point. This indicates that the vHs of $\pi^*_2$ band is located below $E_F$, which corresponds to Fig. 4(f). The calculated electron densities are $n_{Ca,\pi^*1} = 5.5 \times 10^{14}$ cm$^{-2}$ for $\pi^*_1$ and $n_{Ca,\pi^*2} = 1.2 \times 10^{15}$ cm$^{-2}$ for $\pi^*_2$, respectively, and thus the total electron density at $E_F$ is $n_{Ca,\pi^*} = n_{Ca,\pi^*1} + n_{Ca,\pi^*2} = 1.9 \times 10^{15}$ cm$^{-2}$ in the $\pi^*$ bands. The obtained carrier densities are summarized in Table 1.

As the origin of the superconductivity in the Ca-intercalated graphene, the ILB crossing $E_F$ was proposed (3), accordingly we mention its effects on the superconductivity. In Fig. 1(h), in the Ca-intercalated case it is difficult with the photon energy 21.2 eV to conclude whether the ILB exists or not due to the superimposed bands of the SiC bulk at ~ 0.5 eV below the $E_F$ at the Γ point. To see the ILB bands, in Fig. 4(g) the ARPES image with photon energy of 40.8 eV by HeIIα light is displayed, and it shows a parabolic band crossing $E_F$ with its bottom at ~ 0.5 eV below $E_F$ (Fig. 4(h) for clarity). This is the ILB which does not appear for the Li-intercalated graphene. Based on the fact that the ILB overlaps with the inner Dirac cone $\pi^*_1$ near $E_F$, we estimated the electron density of the ILB at $E_F$ to be $n_{Ca,ILB} = 1.9 \times 10^{14}$ cm$^{-2}$. Details of the ILB is discussed in the next section.

**Discussion**

The most notable characteristic in our Ca-intercalated graphene is the quite large carrier (electron) density which is essential for the realization of superconductivity. Let us discuss the origin of it hereafter. In the free-standing C$_6$CaC$_6$ (C$_6$LiC$_6$) model, since the (√3×√3)R30° unit cell contains one Ca (Li) atom, the atomic density of Ca (Li) is equivalent to the density of the (√3×√3)R30° unit cell $N_{\sqrt{3}} = 6.36 \times 10^{14}$ cm$^{-2}$. This is almost equal to the total electron density $n_{Li,tot}$ in the case of the Li-intercalation ($n_{Li,tot} = n_{Li,\pi^*} \sim N_{\sqrt{3}}$), which leads to a conclusion that each of the intercalated monovalent Li atoms provides one electron to the bilayer graphene. On the other hand, in the Ca-intercalated graphene, the total electron density of the $\pi^*$ bands ($n_{Ca,\pi^*}$) and the ILB ($n_{Ca,ILB}$) is $n_{Ca,tot} = 1.9 \times 10^{15}$ cm$^{-2}$, which is ~ 3.5 times as large as that of Li case $n_{Li,tot}$ as shown in Table 1. This value of $n_{Ca,tot}$ is so large compared with the electron density of other 2D superconductors (*ex.* 4.8×10$^{14}$

cm$^{-2}$ in superconducting Tl,Pb on Si(111) with $T_C$ = 2.26 K (35)). In the free-standing C$_6$CaC$_6$ model, even if each of the intercalated divalent Ca atoms donates two electrons, the maximum electron density is $2N_{\sqrt{3}}$ = 1.27×10$^{15}$ cm$^{-2}$, which is not enough to explain the high electron density $n_{Ca,tot}$ observed here. Thus, there has to be another source of electrons that can provide at least $n_{Ca,tot}$ - $2N_{\sqrt{3}}$ = 6.6×10$^{14}$ cm$^{-2}$ of electrons into the graphene layers.

The excess Li and Ca atoms on the topmost surface are not likely to be the additional source of electrons because they form 3D islands providing no carriers as seen in ARPES results (Figs. 1(f,g)). Thus, a probable scenario is that the terminating Li/Ca atoms at the surface of the SiC substrate provide electrons into the graphene layers. Since the carrier density of the π* bands in the Li-intercalated graphene $n_{Li,\pi*}$ is roughly consistent with the atomic density in (√3×√3)R30º superstructure, the terminating Li on the surface of SiC does not provide electrons to graphene. On the other hand, since Ca is divalent and a Ca-Si bond needs only one electron for a Ca atom to bond with a Si atom, it can donate another electron to the graphene layer. Hence, in the rough estimation almost all of the atoms terminating SiC are Ca atoms assuming that the termination periodicity is (√3×√3)R30º because the density of graphene-(√3×√3)R30º (= 6.4×10$^{14}$ cm$^2$) is quite close to $n_{Ca,tot}$ - $2N_{\sqrt{3}}$ (= 6.6×10$^{14}$ cm$^2$).

Therefore, the two significant roles of the interface between graphene and the SiC substrate are clarified now; (1) to change initially monolayer graphene accompanied by a buffer layer into bilayer graphene, which enables Li and Ca atoms to be intercalated between graphene layers, and (2) to supply electrons to the graphene layers by terminating the SiC surface with Ca atoms in addition to electrons doped by the Ca atoms between two graphene layers. These realize the very high electron density inducing the superconductivity.

As mentioned in Figs. 4(g,h), the ILB appears at the Γ point after Ca-intercalation, which is similar to that reported previously in the Ca-intercalated graphene which is initially bilayer above the buffer layer (3). The observed Fermi surface is composed of the π*$_1$, π*$_2$ bands (Fig. 4(b)) and the ILB, which is in good agreement with that of the free-standing C$_6$CaC$_6$ model proposed by Margine et al. (11). In Ref. (11), they theoretically show that in the case of the free-standing C$_6$CaC$_6$ model, both the ILB and the π* bands are required for inducing the superconducting full-gap states at the Fermi surfaces and the ILB has an essential role in enhancing the electron-phonon coupling. In our study, the importance of the ILB for superconductivity is suggested from the fact that the ILB is not occupied in the non-superconducting Li-intercalated graphene while it is occupied in the superconducting Ca-intercalated graphene. On the other hand, our Ca-intercalated bilayer graphene has the dome-like tendency of $T_C$ as shown in Fig. 3(d), which is not a feature of typical s-wave BCS superconductors. In addition, the very high carrier density discussed above for our Ca-intercalated graphene may also indicate a deviation from the free-standing C$_6$CaC$_6$ model.

In our graphene system, we need to think about flat bands having the vHs at the M' point which is theoretically predicted to induce unconventional superconductivity (13-16) similar to the unconventional superconductivity of cuprates. As shown above, in the superconducting Ca-intercalated graphene, $E_F$ is not exactly at $E_{vHs}$ of neither π*$_1$ nor π*$_2$ (~ 0.05 eV beneath the vHs of π*$_1$ and ~ 0.8 eV above the vHs of π*$_2$ as shown in Fig. S4) whereas in the non-superconducting Li-intercalated graphene the vHs of π*$_2$ is quite close to $E_F$. Some previous theoretical studies suggest that unconventional superconductivity can occur in the graphene intercalation compounds even if the $E_F$ is located at ~ ± 0.25 eV from $E_{vHs}$ (13,16). In those theories, the Fermi surface nesting accompanied by a hexagonal-shaped Fermi surface is expected when $E_F$ is near $E_{vHs}$. However, the Fermi surface of π*$_2$ (Fig. 4(a)) in our Li-intercalated graphene shows a circular shape suggesting negligible nesting. Therefore, a sign of the unconventional superconductivity caused by the vHs cannot be seen with ARPES in our Ca-intercalated graphene although some unusual properties such as the significant large carrier density and the "dome-shape" relation of $T_C$ - normal-state-conductivity are observed.

In conclusion, by controlling the atomic structures and evaluating the electronic structures and transport properties, we have clarified the origin of the superconductivity accompanied with the structural transformation at the interface between graphene and the SiC substrate in the Ca-intercalated epitaxial graphene on SiC(0001). With the change of the buffer layer into graphene, an initially monolayer graphene is found to become bilayer graphene, which allows the in-between-intercalation of Li and Ca atoms. Then, we have found that superconductivity is realized only when Ca atoms are intercalated in-between the graphene layers. In addition, the Ca atoms terminating the SiC surface act as electron donors to induce the superconductivity, because the observed total electron density at $E_F$ ($1.9×10^{15}$ cm$^{-2}$) exceeds the electron density expected for a free-standing $C_6CaC_6$ model. Therefore, the contribution from the interface which has not been focused on so far is found to be crucial for the superconductivity in the Ca-intercalated graphene in terms of both atomic structure and electronic states. This work sheds light on the physics of the superconductivity in the intercalated graphene systems, which also suggests the possibility of the superconductivity in graphene intercalated by other metal atoms through controlling the surface termination and intercalation.

**Materials and Methods**

### Graphene fabrication and intercalation

A monolayer graphene was grown on the flat Si-face of a *n*-type 4H-SiC(0001) single crystal. After degassing under the high vacuum condition ($10^{-6}$-$10^{-7}$ Torr), the SiC substrate was heated up to ~1620°C for 20 min under ~1 atm Ar gas. The quality and the number of the graphene layers were confirmed by *ex-situ* Raman spectroscopy and ARPES measurements. The fabricated graphene sample was installed into an ultrahigh vacuum (UHV) chamber of ~$10^{-10}$ Torr for intercalations. Li and Ca atoms were deposited by using a Li dispenser (SAES Getters) and a Knudsen cell made by ourselves, respectively. And the substrate temperature was controlled by the direct-current Joule heating. Further detailed information on the sample fabrication is explained in the result section. In all measurements, the intercalation conditions were in common and the intercalated samples were transferred from the preparation chambers into the respective measurement chambers without exposure to air in order to protect the sample surface structure.

### ARPES measurements

ARPES measurements were performed with a hemispherical analyzer at room temperature. We used two different UV sources of HeIα (21.2 eV) HeIIα (40.8 eV) with energy and angle multi-detections Omicron Scienta R4000 and R3000, respectively. Fig. 4(g) and S4(a,b) are the results of ARPES measurements carried out with HeIIα radiation while the others were measured by using HeIα radiation.

### STM measurements

STM images were obtained by ultralow temperature multi-function scanning probe microscope (ULT-MF-SPM), which is based on the ultralow temperature scanning tunneling microscope setup (41) with a PtIr tip under UHV condition. The sample was kept below 100 mK while measuring its surface topography.

### Transport measurements

*In-situ* transport measurements in ultrahigh vacuum were performed with USM-1300S (UNISOKU) whose original STM probe is replaced by a four-point probe (4PP) containing four copper wires with a diameter of 100 μm (42). The 4PP measurements enable us to obtain an intrinsic sample resistance excluding the effect of the contact resistance. In addition, the dual configuration method eliminates an uncertainty caused by the unintentional changes in the probe distances. The sample temperature was controlled continuously with the lowest limit of ~0.8 K and a magnetic field up to 7 T was applied to the sample surface in the out-of-plane direction.

**Acknowledgments**

We thank H. Fukuyama in Cryogenic Research Center of the University of Tokyo for help with STM measurements. A part of This work was conducted in Research Hub for Advanced Nano Characterization, The University of Tokyo, and NanofabPF, Tokyo Institute of Technology (Grant No. JPMXP09F20IT0008), supported by "Nanotechnology Platform Program" of MEXT, Japan. H.T. was supported by JSPS Research Fellowships for Young Scientists and the MERIT-WINGS program of the University of Tokyo.

**Funding:** This research was partly supported by JSPS KAKENHI Grant Nos. 20H00342, 20H02616, 18H03877, 19K15443, 21K14533, and Grant-in-Aid for JSPS Fellows No. 20J11972.

**Author contributions:** H.T. and Y.E. prepared the samples. H.T., K.H., S.S. and R.H. carried out the transport measurements. H.T., M.H., S.I., T.I., K.H., R.A, T.H. and F.K. performed the ARPES measurements. Y.E. and T.M. conducted the STM measurements. R.A. and S.H. conceived the project. H.T., R.A., S.I. and S.H. wrote the manuscript with input from all authors.

**Competing interests:** The authors declare that they have no competing interests.

**Data and materials availability:** All data needed to evaluate the conclusions in the paper are present in the paper and/or the Supplementary Materials. Additional data related to this paper may be requested from the authors.

**Figures and Tables**

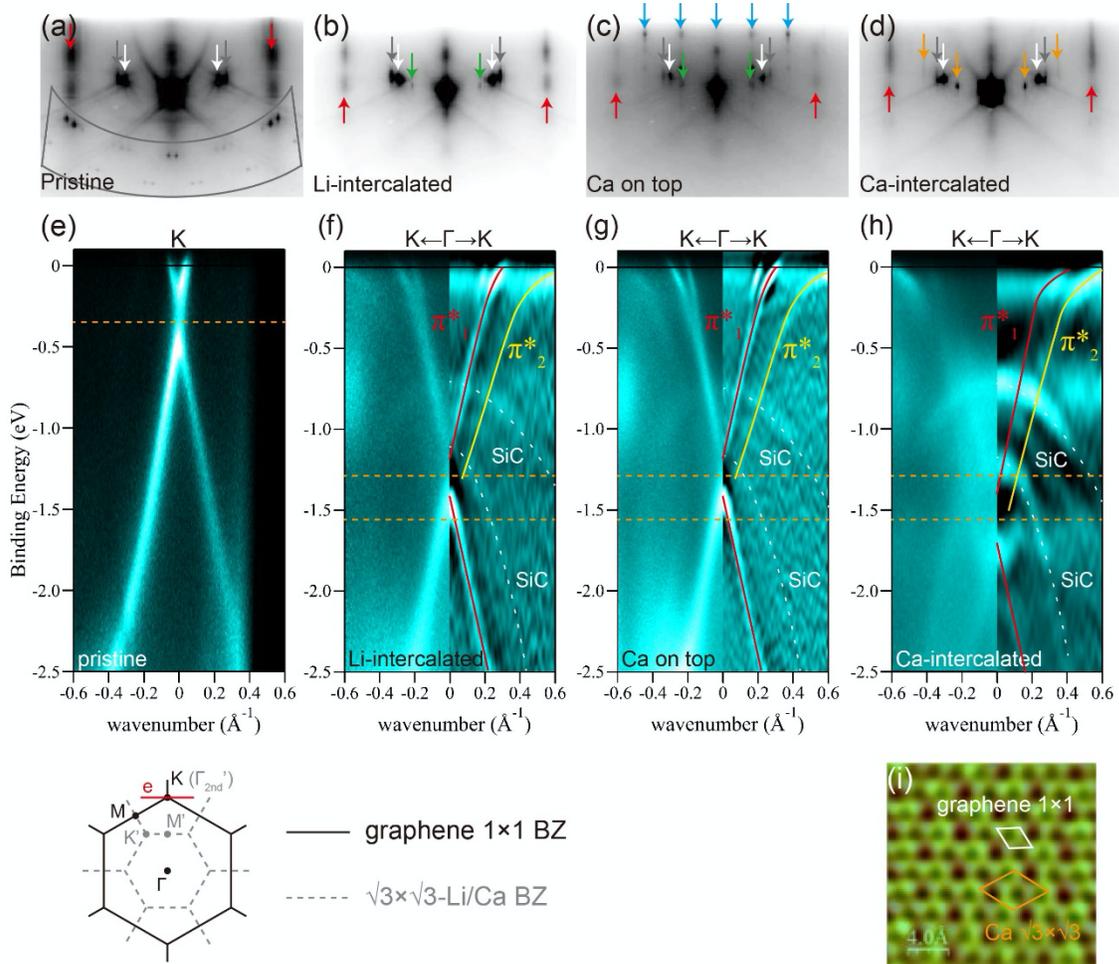

**Fig. 1. RHEED patterns, ARPES results and a STM image in the sample fabrication process of the Ca-intercalated graphene.** (a-d) RHEED patterns from the samples. Each arrow shows RHEED spots of SiC(0001)-1×1(white), graphene-1×1 (red arrows), (6√3×6√3)R30º of the buffer layer (gray arrows and lines surrounding the spots), (√3×√3)R30º-Li (green arrows), bulk-like Ca layer (light blue arrows) and (√3×√3)R30º-Ca (orange arrows). (e-h) Band dispersions measured by ARPES. The right half of each image is a second derivative one. The horizontal white dashed straight lines indicate the energy level of the Dirac points. Brillouin zones of graphene and (√3×√3)R30º-Li/Ca are indicated by solid and dashed lines, respectively. (e) is measured around the K point along a line perpendicular to the Γ-K-M direction (red line), whereas others are measured around the Γ point along the K-Γ-K direction. (a,e) Pristine monolayer graphene. (b,f) Li-intercalated graphene after Li deposition. (c,g) after Ca-deposition on the Li-intercalated graphene. (d,h) Ca-intercalated graphene after annealing. (i) A filtered STM image of the Ca-intercalated graphene with a size of 2×2 nm². The white and orange diamonds indicate unit cells of graphene-1×1 and (√3×√3)R30º-Ca, respectively.

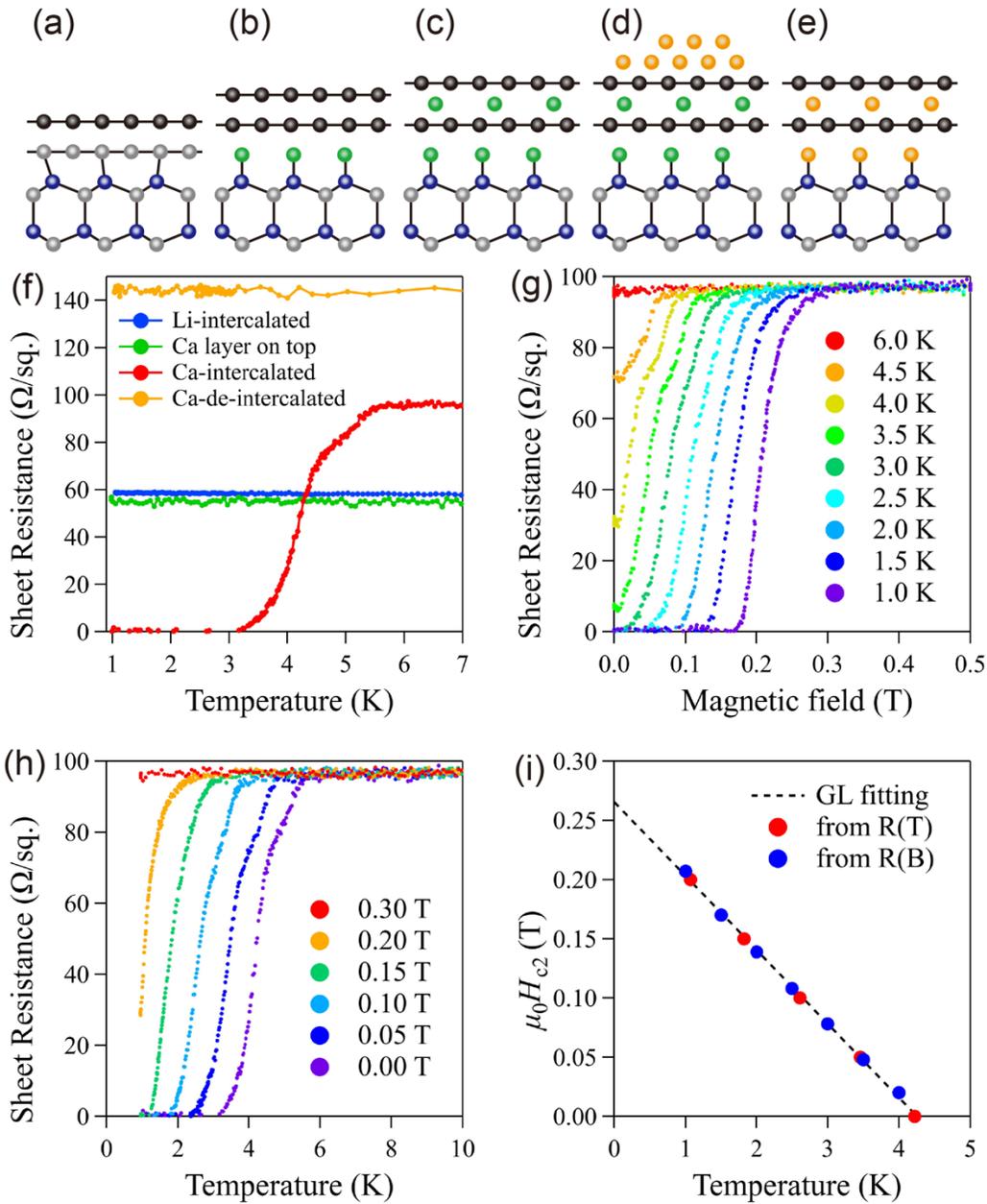

**Fig. 2. Structural model and transport properties of Ca-intercalated graphene.**
(a-e) structural models during the sample fabrication process suggested from Fig. 1. (a) Pristine graphene. (b) Quasi-freestanding bilayer graphene on the Li-terminated SiC. (c) Li-intercalated graphene after further Li deposition on (b). (d) after Ca deposition on (c). (e) Ca-intercalated graphene after annealing (d). (f) Temperature-dependent sheet resistance $R(T)$ of Li-intercalated (blue), Ca-deposited (green), Ca-intercalated (red) and Ca-de-intercalated (orange) graphene. (g) Sheet resistance as a function of perpendicular-to-the-plane magnetic field at constant temperatures varied from 1.0 K to 6.0 K. (h) $R(T)$ measured under constant magnetic field up to 0.30 T. (i) Upper critical magnetic field extracted from (g) and (h) indicated by blue and red dots, respectively. The dashed line is the result of Ginzburg-Landau fitting.

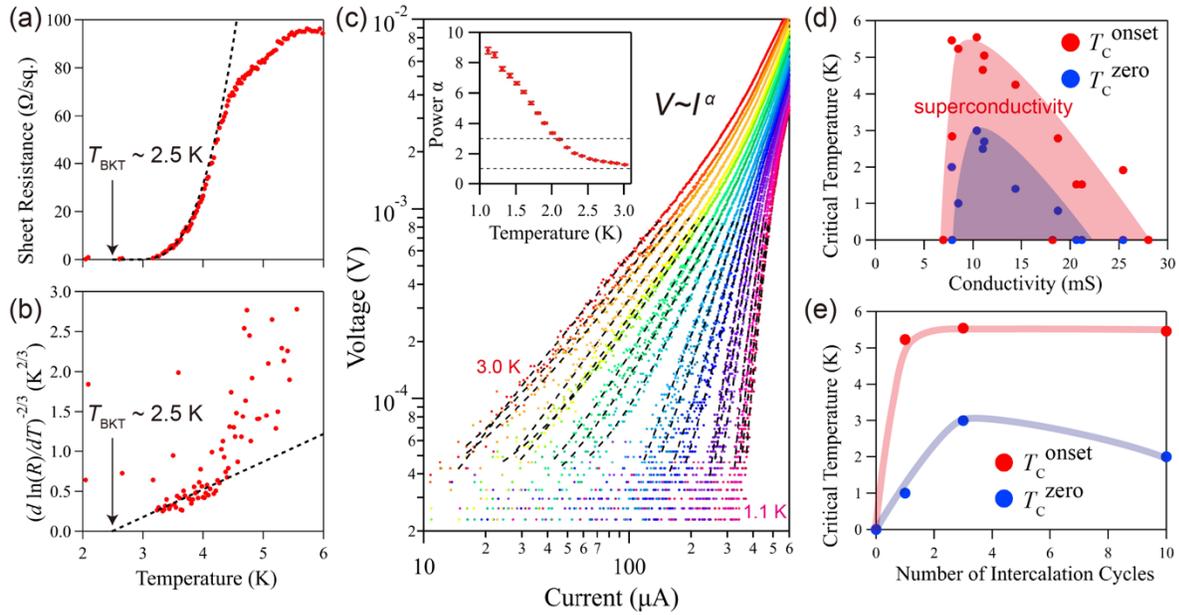

**Fig. 3. Observation of BKT transition and the details of critical temperature $T_C$.** (a) Fitting by the formula of the BKT transition (Eq. 2 in the main text) with the data of the red curve in Fig. 2(f). The black dashed line indicates the fitting result. (b) Replot of (a) with a vertical axis of $(d \ln(R(T))/dT)^{-2/3}$. (c) Voltage-current curves from 1.1 K to 3.0 K plotted in double logarithmic scale. Fitting results by the relation $V \sim I^\alpha$ are shown with dashed lines. Inset: power $\alpha$ as a function of temperature. (d) Critical temperature $T_C$ as a function of normal-state conductivity (inversion of normal-state resistivity). $T_C^{onset}$ (red) and $T_C^{zero}$ (blue) are the temperatures where $R(T)$ starts decreasing and becomes zero, respectively. (e) Relation between $T_C$ and the number of intercalation cycles. One cycle is determined by a series of Li-intercalation, Ca-deposition and annealing.

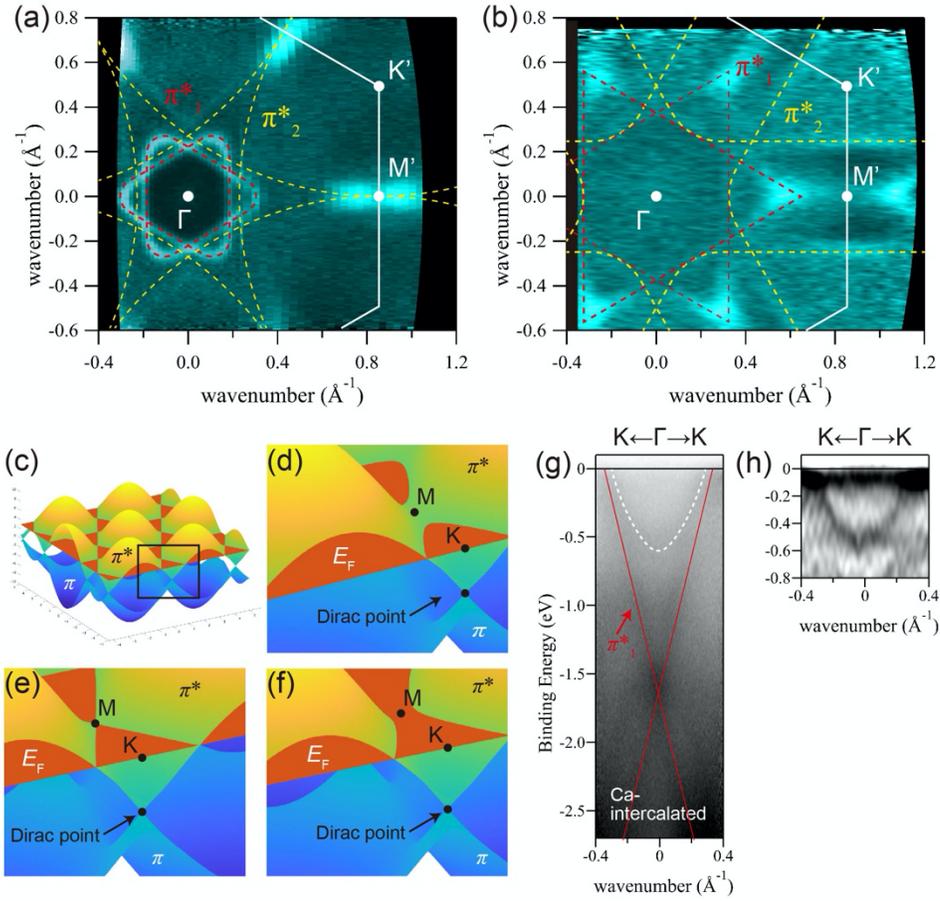

**Fig. 4. Fermi surface mapping and band dispersion measured by HeIIα light source.**
(a,b) Fermi surface mapping of Li- and Ca-intercalated graphene, respectively. Solid white lines indicate the first Brillouin zone of (√3×√3)R30º superstructure. (c-f) 3D schematic images of band dispersion with different Fermi level (red plane). The Dirac point at the K point is indicated by a black arrow and vHs is located at the M point. Due to the folding of graphene bands into the √3×√3 BZ, the Dirac point and vHs appear at the Γ' and M' point in (a,b), respectively. (c) Calculated band structure of a free-standing monolayer graphene. Yellow and blue curved surfaces are energy bands of graphene and the red plane indicates $E_F$. The area surrounded by a black rectangle is zoomed in (c-e) with different Fermi levels. (d) $E_F$ is below vHs. (e) $E_F$ is at the same energy level as vHs. (f) $E_F$ is above vHs. (g) Band dispersion of Ca-intercalated graphene around the Γ point along the the K-Γ-K direction measured by HeIIα light source (40.8 eV). The ILB is indicated by a dashed white line. (h) Second derivative of (g) near the Fermi level showing the ILB.

|  | $\pi^*_1$ [×$10^{14}$ cm$^{-2}$] | $\pi^*_2$ [×$10^{14}$ cm$^{-2}$] | ILB [×$10^{14}$ cm$^{-2}$] | Total [×$10^{14}$ cm$^{-2}$] |
|---|---|---|---|---|
| Pristine | 0.11 | - | - | 0.11 |
| Li-intercalated | 1.8 | 3.6 | - | 5.3 |
| Ca-intercalated | 5.5 | 11.9 | 1.9 | 19.3 |

**Table 1. Summarized carrier density of the pristine, Li-intercalated and Ca-intercalated graphene**

# Supplementary Materials for

## Two-dimensional superconductivity of the Ca-intercalated graphene on SiC: vital role of the interface between monolayer graphene and the substrate

Haruko Toyama*, Ryota Akiyama*, Satoru Ichinokura, Mizuki Hashizume,
Takushi Iimori, Yukihiro Endo, Rei Hobara, Tomohiro Matsui, Kentaro Horii,
Shunsuke Sato, Toru Hirahara, Fumio Komori, Shuji Hasegawa

*Corresponding author. Email:
h.toyama@surface.phys.s.u-tokyo.ac.jp
akiyama@surface.phys.s.u-tokyo.ac.jp

**This PDF file includes:**

Supplementary Text
Figs. S1 to S4

**Error estimation of temperature at current-voltage measurements in Fig. 3(c)**

We took into account the deviation of the observed temperature from the actual one during the *I-V* measurements which are shown in Fig. 3(c). This is because a current high enough to break the superconductivity (~ 700 μA) is applied repeatedly to observe the BKT transition, and the actual sample temperature increases inevitably by the Joule heating. We suggest that this is the reason why $\alpha$ becomes 3 at $T$ = 2.1 K, which is slightly lower than $T_{BKT}$ = 2.5 K obtained from Fig. 3(a,b) by Halperin-Nelson's theory.

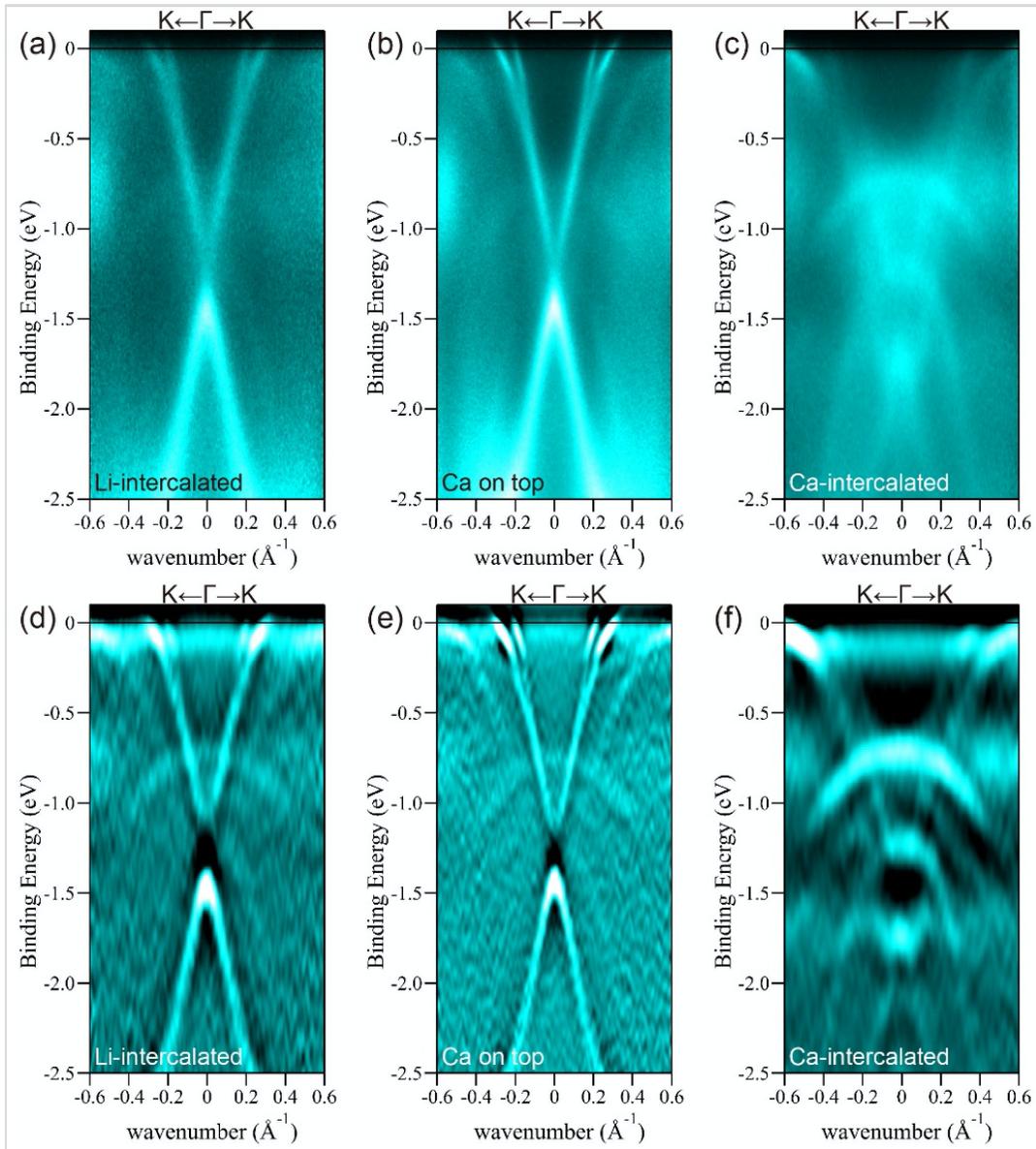

**Fig. S1. ARPES images of Fig. 1(f-h) without guide lines.**

(a-c) ARPES images of Fig. 1(f-h) without guide lines. (d-f) second derivative images of S1(a-c), respectively. (a,d) Li-intercalated graphene. (b,e) after Ca-deposition on Li-intercalated graphene. (c,f) Ca-intercalated graphene after annealing.

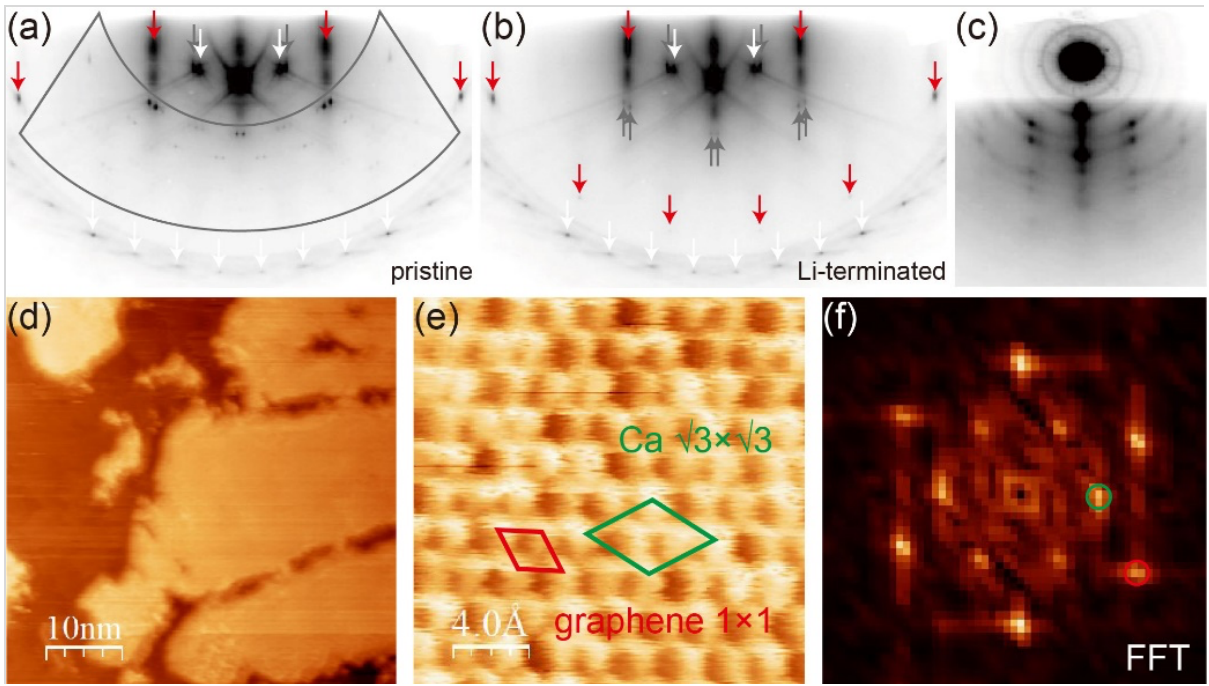

**Fig. S2. Additional RHEED patterns and STM images**

(a,b) RHEED patterns of the pristine graphene and the Li-terminated graphene before starting the Li-intercalation. Most of (6√3×6√3)R30º spots from the buffer layer (gray arrows and lines) disappear suggesting the decoupling of it from the SiC substrate. At the same time the graphene 1×1 spots become stronger due to the formation of additional quasi-freestanding graphene layer. (c) RHEED pattern of the Ca-intercalated graphene with a small incident angle. The intensity of these 3D spots is enhanced with a smaller incident angle suggesting that some 3D-like islands structures exist on the topmost surface. (d) Topographic STM image of a larger area (50×50 nm$^2$). The bright areas are expected to be the Ca-intercalated graphene while the dark areas are not intercalated. (e) Enlarged image of the bright area in (d) before the filtering. The red and green meshes indicate a unit cell of graphene-1×1 and (√3×√3)R30º-Ca, respectively. (f) Fast Fourier transformed (FFT) image of S1(e) showing the periodicities of graphene-1×1 and (√3×√3)R30º-Ca indicated by red and green circles, respectively.

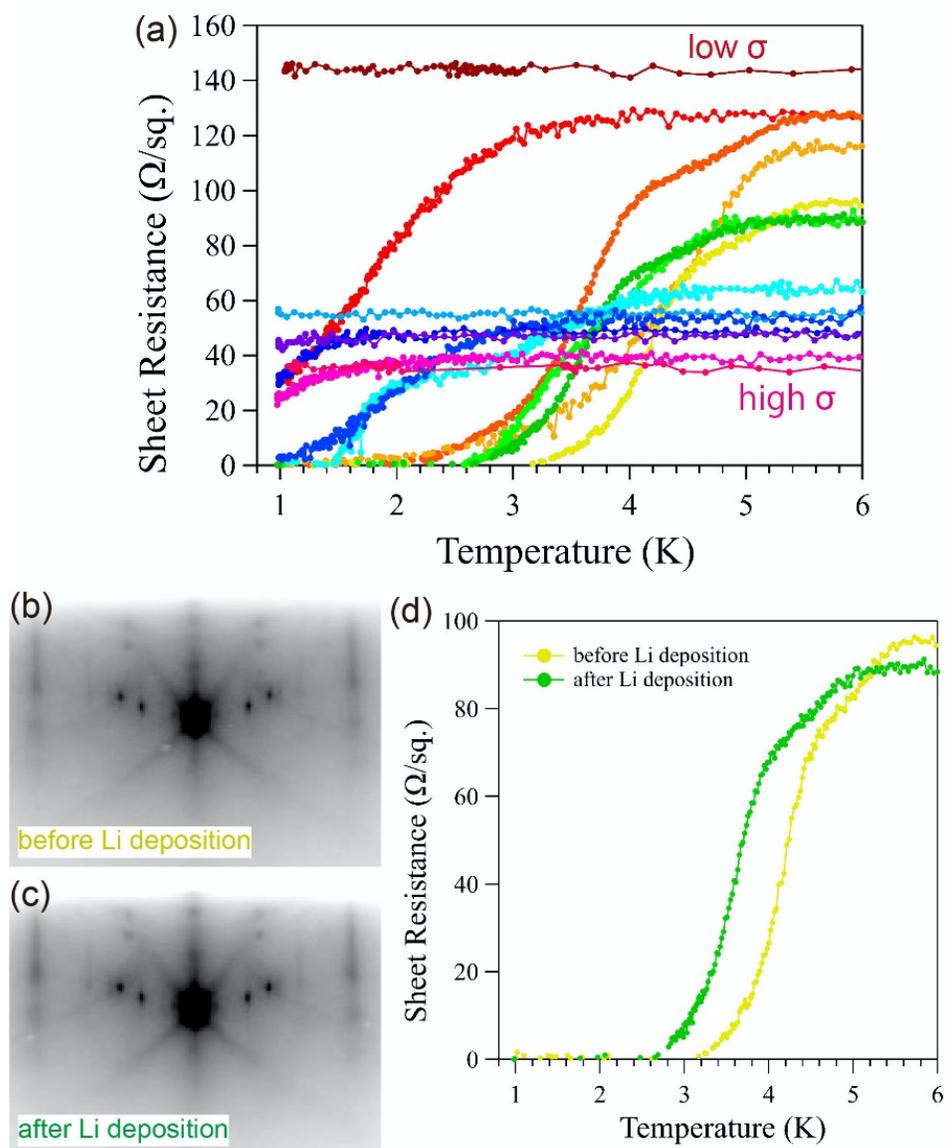

**Fig. S3. Original data of electrical transport measurements in Fig. 3(d)**
(a) The sheet resistance of 14 different samples of the Ca-intercalated graphene as a function of the temperature. The normal-state resistance of each sample was controlled by changing the sample growth conditions (the deposition amount of Li and Ca atoms and the annealing temperature) (b,c) RHEED patterns of a superconducting Ca-intercalated graphene sample before and after the additional Li-deposition to induce the carrier-transfer further into graphene. (d) Temperature dependences of the sheet resistance of samples shown in (b,c).

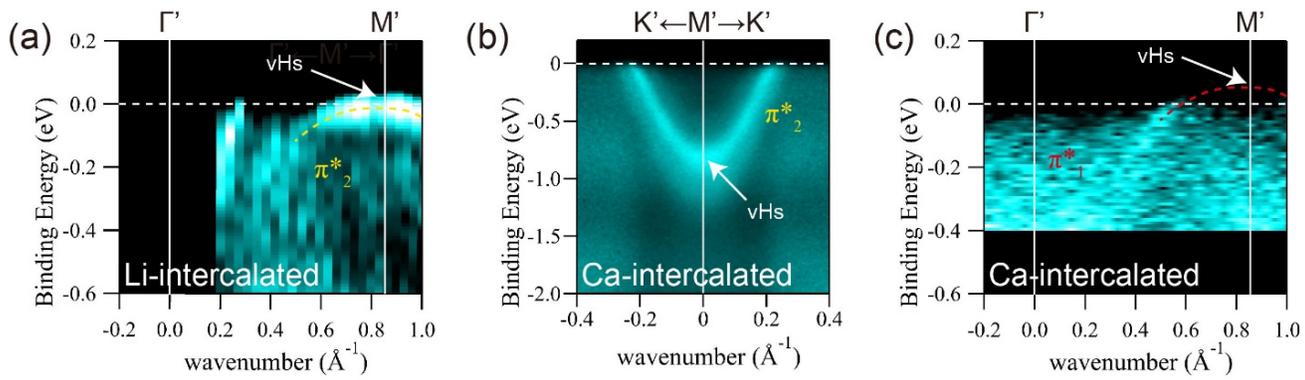

**Fig. S4. Positions of vHs in Li- or Ca-intercalated graphene**

(a) Band dispersion of Li-intercalated graphene along the Γ'-M' direction. The yellow dashed line is a guide line for the $\pi^*_2$ band around the M' point. (b) Band dispersion of Ca-intercalated graphene around the M' point along the K'-M'-K' direction. (c) Band dispersion of Ca-intercalated graphene along the Γ'-M' direction. The red dashed line is a guide line for the $\pi^*_1$ band with assuming that its shape is the same as that of the $\pi^*_2$ band of Li-intercalated graphene shown in (a). (a-c) Expected positions of the vHs are indicated by arrows.